# Experimental demonstration of a graph state quantum error-correction code

B. A. Bell[1], D. A. Herrera-Martí[2], M. S. Tame[3,4*], D. Markham[5], W. J. Wadsworth[6] & J. G. Rarity[1*]

[1] *Centre for Communications Research, Department of Electrical and Electronic Engineering, University of Bristol, Merchant Venturers Building, Woodland Road, Bristol, BS8 1UB, United Kingdom*

[2] *Centre for Quantum Technologies, National University of Singapore, 3 Science Drive 2, Singapore 117543, Singapore*

[3] *School of Chemistry and Physics, University of KwaZulu-Natal, Durban 4001, South Africa*

[4] *National Institute for Theoretical Physics, University of KwaZulu-Natal, Durban 4001, South Africa*

[5] *CNRS LTCI, Département Informatique et Réseaux, Telecom ParisTech, 23 avenue d'Italie, CS 51327, 75214 Paris CEDEX 13, France*

[6] *Centre for Photonics and Photonic Materials, Department of Physics, University of Bath, Claverton Down, Bath, BA2 7AY, United Kingdom*

**Scalable quantum computing and communication requires the protection of quantum information from the detrimental effects of decoherence and noise. Previous work tackling this problem has relied on the original circuit model for quantum computing. However, recently a family of entangled resources known as graph states has emerged as a versatile alternative for protecting quantum information. Depending on the graph's structure, errors can be detected and corrected in an efficient way using measurement-based techniques. In this Letter we report an experimental demonstration of error correction using a**

**graph state code. We have used an all-optical setup to encode quantum information into photons representing a four-qubit graph state. We are able to reliably detect errors and correct against qubit loss. The graph we have realized is setup independent, thus it could be employed in other physical settings. Our results show that graph state codes are a promising approach for achieving scalable quantum information processing.**

# Introduction

Quantum error correcting codes (QECCs) constitute fundamental building blocks in the design of quantum computer architectures[1]. It was realized early on that using QECCs[2-6] to counteract the effects of decoherence and noise provides a means to increase the coherence time of the encoded information. This enhancement is crucial for enabling a range of speedups in quantum algorithms. Here, the threshold theorem[7] ensures that a quantum computer built with faulty, unreliable components can still be used reliably to implement quantum tasks using QECC techniques[8,9], so long as the noise affecting its parts is below a given threshold. A great deal of effort is currently being invested in designing new quantum codes to increase the threshold. In this context a computational paradigm especially well suited for quantum error correction is measurement-based quantum computation[10-12] (MBQC), in which a resource state consisting of many entangled qubits is prepared before the computation starts. In MBQC, an algorithm is enacted by performing sequential measurements on the resource state in such a way that the output of the computation is stored in the unmeasured qubits. Photonic technologies[13] have enjoyed enormous success in the generation of a variety of resource states for MBQC[14-18] and in the implementation of computational primitives[19-32]. Importantly, QECCs can be embedded in the resource states for MBQC in several inequivalent ways[33-35], and of particular theoretical interest, due to their large thresholds, are the topological QECC embeddings[36-42]. However, while there has been an experimental proof-of-principle for topological encoding[43], overall these codes remain largely out of reach of current technologies due to the size and complexity of the resources required. An alternative and more compact approach is offered by the theory of graph codes[44-47], where very general QECCs can be used within the framework of MBQC to account for different noise scenarios. Graph codes are based on the stabilizer formalism and are thus relevant for both MBQC and the original circuit model.

In this work we report the experimental demonstration of a quantum error-correcting graph code. We have used an all-optical setup to encode quantum information into photons representing the code. The experiment was carried out for the smallest graph code capable of detecting one quantum error, namely the four-qubit code[51-53] [[4,1,2]]. Here, [[n,k,d]] is the standard notation for QECCs, where $n$ denotes the number of physical qubits, $k$ is the number of logical qubits encoded and $d$ is the distance, which indicates how many errors can be tolerated and depends on information about the error: a code of distance $d$ can correct up to $\lfloor(d-1)/2\rfloor$ arbitrary errors at unspecified locations. On the other hand, if we know where the error occurs the code can correct up to $d-1$ errors (equivalently erasures or loss errors), or it can detect up to $d-1$ errors at unspecified locations (without necessarily being able to correct them). The four-qubit code used in our experiment has a distance of $d=2$, so it can correct up to one quantum error or a loss error at a *known* location and can detect up to one quantum error at an *unknown* location. This has applications in several key areas of quantum technologies besides the obvious goal of fault-tolerance[54-57], for example in communication over lossy channels, lossy interferometry and secret sharing. While the four-qubit code has been realized before[50], the work was restricted to quantum error-correction in the original circuit model. Here we go beyond this and show how to realize the code using an entangled graph state in the promising context of MBQC and fully characterize its performance. One of the important distinctions of our work is that the graph state resource for the code is generated first and the quantum information is then teleported into it, following closely the model for MBQC. We show that by measuring an external ancilla qubit its information can be transferred into the logical subspace of the code, which after undergoing a noisy channel can be decoded to retrieve the original information with high quality. In addition, recent work incorporating quantum error-correction using a measurement-based approach has considered basic protection against phase errors, where the location of the error is known[58,59]. In this work we lift these restrictions and experimentally demonstrate a graph code that can be used within the MBQC framework to provide protection against arbitrary general quantum errors and loss, where the location of the error is known, as well as the detection of general quantum errors where the location is unknown. We have successfully demonstrated all elements of error correction in our experiment, including in sequence the encoding, detection and correction of errors, and we have verified the quality of each of these steps separately. In general, the versatility of graph codes, such as the one we have demonstrated, can further be increased by generalizing them to codeword-stabilized (CWS) codes[60], where a given graph is supplemented with a (possibly non-additive) classical code that corrects the classical errors induced by the stabilizer structure. The theory of CWS codes is the most general theory of QECCs to date, as it encompasses graph codes, of which the four-qubit graph code we have realized is the simplest instance, and non-additive codes. Thus, the graph encoding we report is amenable to be used with the more general CWS codes. Demonstrations of compact QECC schemes, such as the one we have performed, are of the utmost importance to the design and characterization of noise protection in a number of different physical architectures at present. They constitute the necessary first steps towards large-scale quantum computers. Our experimental demonstration and its full analysis contribute to helping achieve these first steps.

# Results

The resource state we used for demonstrating the four-qubit graph code was generated as shown in Figure 1a using two photonic crystal fibre (PCF) sources[61,62,63] which each produce correlated pairs of photons via spontaneous four-wave mixing when pumped by picosecond laser pulses. One of the sources was in a Sagnac loop configuration, such that the PCF is pumped in both directions, with one direction producing horizontally polarized signal-idler pairs, $|H\rangle_{i_2} |H\rangle_{s_2}$, and the other producing vertical pairs, $|V\rangle_{i_2} |V\rangle_{s_2}$. When the two paths are combined at a polarizing beam-splitter (PBS), the polarizations of a pair outside the loop are entangled in a Bell state (see Methods), $\frac{1}{\sqrt{2}}(|H\rangle_{i_2} |H\rangle_{s_2} + |V\rangle_{i_2} |V\rangle_{s_2})$. The other source is used to produce a heralded signal photon in the state $|+\rangle_{s_1}$, where $|\pm\rangle = \frac{1}{\sqrt{2}}(|H\rangle \pm |V\rangle)$. This is overlapped with the signal photon from the entangled pair at a PBS, performing a post-selected fusion operation[64,65,66]. Conditioned on a four-fold coincidence detection this will leave a GHZ state on three of the photons, $\frac{1}{\sqrt{2}}(|H\rangle_{s_1}|H\rangle_{i_2} |H\rangle_{s_2} + |V\rangle_{s_1}|V\rangle_{i_2} |V\rangle_{s_2})$. This state is equivalent to a three-qubit linear cluster state up to local rotations, which are applied to the end qubits using half-wave plates (HWPs) on the two signal modes to give $\frac{1}{\sqrt{2}}(|+\rangle_{s_1}|H\rangle_{i_2} |+\rangle_{s_2} + |-\rangle_{s_1}|V\rangle_{i_2} |-\rangle_{s_2})$. Additional path degrees of freedom are then used to expand the state into a five-qubit linear cluster[18]. Here, the signal photons are each split into two paths using PBSs, so that the path they take is correlated with their polarization, and the transmitted and reflected paths, $p_1$ and $p_2$, are labelled as $|0\rangle$ and $|1\rangle$ for the additional qubits. To detect a path qubit in a particular basis, the paths are recombined at a 50:50 beam-splitter (BS), which performs a Hadamard rotation on the path, independent of the polarization. By shifting the relative phase $\theta$ before this, using tilted glass plates, the path qubit can be detected after the BS in any state in the equatorial plane of the Bloch sphere given by $\frac{1}{\sqrt{2}}(|0\rangle + e^{i\theta}|1\rangle)$, e.g. the Pauli X and Y bases. Measurements in the Pauli Z basis can be achieved by blocking one or the other interferometer path, in which case the BS reduces the measurement rate by

50%. The polarization qubits are measured after the path qubits using a quarter-wave plate (QWP), HWP and PBS chain[63], followed by a detection of the photon using single-photon avalanche photodiodes. This allows us to measure in the X, Y and Z basis[67]. The state after a path qubit is added to each signal photon, with Hadamard rotations applied to the signal polarizations using a HWP in each path, can be written as

$$|\psi_{lin}\rangle = \frac{1}{2\sqrt{2}} [\ (|+\rangle|0\rangle + |-\rangle|1\rangle)|0\rangle(|0\rangle|+\rangle + |1\rangle|-\rangle) + \quad (1)$$

$$(|+\rangle|0\rangle - |-\rangle|1\rangle)|1\rangle(|0\rangle|+\rangle - |1\rangle|-\rangle)\ ]_{12345}\ ,$$

Which is the five-qubit linear cluster state shown in Figure 1b, where the polarization of photon $s_1$ represents qubit 1 ($|H/V\rangle \to |0/1\rangle$) and its path represents qubit 2 ($|p_1/p_2\rangle \to |0/1\rangle$). Similarly for photon $s_2$, whose polarization represents qubit 5 and its path qubit 4. The polarization of photon $i_2$ represents qubit 3.

The linear cluster state is then transformed into the resource state consisting of the graph code plus ancilla qubit according to the local complementation rules for graph states, as shown in Figure 1b and described in the caption. The resulting graph state can be written compactly as $\frac{1}{\sqrt{2}}(|0\rangle_3|+_L\rangle + |1\rangle_3|-_L\rangle)$, where the logical states of the four-qubit graph code are given by $|0_L\rangle = \frac{1}{\sqrt{2}}(|\phi^-\rangle_{15}|\phi^-\rangle_{42} - |\psi^-\rangle_{15}|\psi^-\rangle_{42})$ and $|1_L\rangle = \frac{1}{\sqrt{2}}(|\psi^+\rangle_{15}|\phi^+\rangle_{42} + |\phi^+\rangle_{15}|\psi^+\rangle_{42})$, with the Bell states given by $|\phi^\pm\rangle = \frac{1}{\sqrt{2}}(|0\rangle|0\rangle \pm |1\rangle|1\rangle)$ and $|\psi^\pm\rangle = \frac{1}{\sqrt{2}}(|0\rangle|1\rangle \pm |1\rangle|0\rangle)$. Here the logical Pauli operators on the codespace are $\overline{X} = Z_1 Z_2 X_4 I_5$ and $\overline{Z} = Z_1 Z_2 Z_4 Z_5$, with $\overline{Y} = i\overline{XZ}$ (see Methods). The total resource state can be written more explicitly as

$$|\psi_{res}\rangle = \frac{1}{2\sqrt{2}} [\ (|+\rangle|+\rangle + i|-\rangle|-\rangle)|-_y\rangle(|+\rangle|+\rangle + i|-\rangle|-\rangle) +$$

$$i(|+\rangle|+\rangle - i|-\rangle|-\rangle)|+_y\rangle(|+\rangle|+\rangle - i|-\rangle|-\rangle)\ ]_{12345}\ ,$$

where the states $|\pm_y\rangle = \frac{1}{\sqrt{2}}(|0\rangle \pm i|1\rangle)$ are the Y eigenstates. To obtain this state from Equation (1) in our experiment, a QWP on idler mode $i_2$ carries out the

required rotation for qubit 3. The transformations for the signal path qubits are implemented by a relabeling of the $|0\rangle$ and $|1\rangle$ paths to $|+\rangle$ and $|-\rangle$, and $\pi/2$ phase-shifts using tilted glass plates. To check the entanglement of the resource, we use an entanglement witness as described in ref. 69. Here, for any GHZ or linear cluster state, it is possible to detect genuine multipartite entanglement (GME) using correlations taken from just two local measurement bases. Since the resource is locally equivalent to a linear cluster state, making corresponding changes to the reference frames of the measurements provides an appropriate witness. The measurements are $X_1X_2X_3X_4X_5$ and $Z_1Y_2Y_3Y_4Z_5$ (see Methods), which result in a witness value of

$$\langle \hat{\mathcal{W}} \rangle = -0.15 \pm 0.03,$$

where the error has been calculated using a Monte Carlo method with Poissonian noise on the count statistics[67]. The negative value of the witness indicates the presence of GME, confirming that all qubits are involved in the generation of the resource. The individual expectation values forming the expression for the witness are shown in Figure 1c. The above witness also sets a lower bound on the fidelity of the state to the ideal case as $F > 0.58 \pm 0.03$.

In order to check the persistency of entanglement in the resource we measure the ancilla qubit using a Z measurement, thus removing it from the graph. For the case that the state $|0\rangle_3$ is measured, the remaining four qubits should be left in the logical code state $|+_L\rangle$, which corresponds to a 'box' cluster-state,

$$|+_L\rangle = \tfrac{1}{2}(|+\rangle|+\rangle|0\rangle|0\rangle + |+\rangle|+\rangle|1\rangle|1\rangle + |-\rangle|-\rangle|0\rangle|1\rangle + |-\rangle|-\rangle|1\rangle|0\rangle)_{1245}.$$

Using the relevant witness in ref. 69 (see Methods) we find the value

$$\langle \hat{\mathcal{W}} \rangle = -0.16 \pm 0.03,$$

showing GME persists even when the ancilla qubit is removed, with $F > 0.58 \pm 0.03$, consistent with the quality of the inital graph state.

In order to encode an arbitrary ancilla qubit $\alpha|0\rangle_3 + \beta|1\rangle_3$ into the four-qubit graph code we measure it in the X basis as shown in Figure 2a. This is a basic quantum information transfer primitive in MBQC and propagates the qubit into the code while at the same time applying a Hadamard operation, so that the qubit is encoded in the Hadamard basis, *i.e.* $\alpha|0\rangle_3|+_L\rangle + \beta|1\rangle_3|-_L\rangle \rightarrow \overline{X}^{s_3}(\alpha|+_L\rangle + \beta|-_L\rangle)$. Thus the encoding of an arbitrary state can be carried out up to a logical byproduct operation $\overline{X}^{s_3}$ depending on the ancilla's measurement result, $s_3 = (0,1)$. Alternatively, an unknown qubit could be entangled with the ancilla qubit via a controlled-phase operation, $C_Z = \text{diag}(1,1,1,-1)$, after which both the unknown state and ancilla are measured in the X basis, transferring the quantum information into the code in the computational basis. We start our characterization of the graph code's performance by analysing the quality of the logical encodings for general input states. To do this we encode the probe states $|0\rangle$, $|1\rangle$, $|+\rangle$ and $|+_y\rangle$ onto the ancilla qubit and measure it in the X basis, as shown in Figure 2b. This is sufficient to reconstruct the encoding process completely as a quantum channel and fully characterise its quality.

The probe state $|0\rangle$ is encoded onto the ancilla qubit using a polarizer in the idler mode $i_2$ with the qubit then measured in the X basis. This propagates the probe state into the code as the state $|+_L\rangle$. This is the box cluster state, which we find to have a witness value of $\langle \widehat{\mathcal{W}} \rangle = -0.11 \pm 0.02$. For convenience we have taken the case where no byproduct is produced during the encoding measurement, *i.e.* $s_3 = 0$. The density matrix for the encoded logical state is shown in Figure 2b and is obtained by measuring in the collective $\overline{X}$, $\overline{Y}$ and $\overline{Z}$ bases of the code, corresponding to local measurements of the four qubits. The fidelity with respect to the ideal case is $F = 0.78 \pm 0.01$. Similarly, using a polarizer in the idler mode we encode the $|1\rangle$ probe state which is propagated into the graph code as $|-_L\rangle$, a state equivalent to the box cluster up to Z rotations on each physical qubit, as $|+_L\rangle = \overline{Z}|-_L\rangle$. A witness value for GME is found to be $\langle\widehat{\mathcal{W}}\rangle = -0.10 \pm 0.03$. The density matrix for this logical state is shown in Figure 2b and the fidelity with respect to the ideal case is $F = 0.77 \pm 0.01$

For the $|+\rangle$ probe state we find that it is naturally encoded into the ancilla qubit in the total graph resource and upon measuring it we expect the logical state $|0_L\rangle$ to be encoded into the four-qubit graph. For the physical qubits this logical state can be rewritten as a rotated GHZ state, $\frac{1}{\sqrt{2}}(|+\rangle|-\rangle|-\rangle|+\rangle + |-\rangle|+\rangle|+\rangle|-\rangle)$. Using a GME witness with two measurement settings (see Methods) we find $\langle\widehat{\mathcal{W}}\rangle = -0.16 \pm 0.03$. The logical density matrix is shown in Figure 2b and the fidelity with respect to the ideal case is $F = 0.78 \pm 0.01$. Finally, for the $|+_y\rangle$ probe state we use a QWP in idler mode $i_2$ and expect the logical state $|-_{y,L}\rangle$ to be encoded into the graph. The $|\pm_{y,L}\rangle$ states are the only encodings not expected to show GME under ideal conditions; instead they are biseparable and composed of two maximally entangled pairs $\frac{1}{\sqrt{2}}(|+\rangle|+\rangle + i|-\rangle|-\rangle)_{12}$ and $\frac{1}{\sqrt{2}}(|+\rangle|+\rangle + i|-\rangle|-\rangle)_{45}$. For the state $|-_{y,L}\rangle$ we find an entanglement witness value for qubit pair (1,2) of $\langle\widehat{\mathcal{W}}\rangle = -0.83 \pm 0.01$ and for pair (4,5) a value of $\langle\widehat{\mathcal{W}}\rangle = -0.70 \pm 0.02$. The logical density matrix is shown in Figure 2b and the fidelity with respect to the ideal case is $F = 0.88 \pm 0.01$.

Using the logical density matrices for the encoded probe states we are able to reconstruct the encoding process as a quantum channel using quantum process tomography[28]. In this case, the encoding transforms a single-qubit input state $\rho$ for the ancilla into the output density matrix $\varepsilon(\rho)$ in the graph code's logical qubit basis and can be formally written as $\rho \rightarrow \varepsilon(\rho) = \sum_{ij} \chi_{ij} M_i \rho M_j^\dagger$. Here, the operators $M_i$ correspond to a complete basis for the Hilbert space allowing any physical channel to be described. We choose the Pauli basis, $M_i = (I, X, Y, Z)$, for the operators so that the elements of the $\chi$ matrix define the channel completely. This allows us to determine the effect of the MBQC information transfer process on the original qubit due to imperfections in the experimental graph resource. In Figure 2c we show the original Bloch sphere for arbitrary input ancilla states and the final reconstructed encoded Bloch sphere using the experimentally determined values from the $\chi$ matrix. The Bloch sphere is reduced slightly in diameter, but overall the structure closely resembles that of the original input states rotated by a Hadamard operation. The process fidelity for the encoding quantifies how close the experiment is to the ideal case and is given by $F_p = \frac{\text{Tr}(\chi_{exp}\chi_{ideal})}{\text{Tr}(\chi_{exp})\text{Tr}(\chi_{ideal})}$, where $\chi_{exp}$ describes the experimental channel and $\chi_{ideal}$ corresponds to a Hadamard rotation. From the channel reconstruction we find $F_p = 0.70 \pm 0.01$.

With the logical encodings characterised we now analyse the performance of the graph code for providing protection against the loss of any of the qubits when the location of the loss is known. In order to see how the graph code tolerates loss, consider the case in which qubit 4 is lost, as shown in Figure 3a. Due to the symmetry of the state, any other qubit can be considered to be lost, with the same recovery procedure applied upon an appropriate rotation of the labelling of the qubits. In the case that we lose qubit 4, the state of the remaining three qubits is found by tracing it out. From the initial state $\alpha|+_L\rangle + \beta|-_L\rangle$ one finds the state $\rho_{251} = \frac{1}{2}(|\phi\rangle\langle\phi| + |\phi^\perp\rangle\langle\phi^\perp|)$, where $|\phi\rangle = \alpha'(|0\rangle|\phi^-\rangle + |1\rangle|\psi^-\rangle) + \beta'(|0\rangle|\psi^+\rangle + |1\rangle|\phi^+\rangle)$ and $|\phi^\perp\rangle = -\alpha'(|1\rangle|\phi^-\rangle + |0\rangle|\psi^-\rangle) + \beta'(|1\rangle|\psi^+\rangle + |0\rangle|\phi^+\rangle)$. Here, the coefficients are $\alpha' = \frac{1}{\sqrt{2}}(\alpha + \beta)$ and $\beta' = \frac{1}{\sqrt{2}}(\alpha - \beta)$. By measuring qubit 2 in the Z basis, we obtain the state $\rho_{51} = \frac{1}{2}(|\varphi\rangle\langle\varphi| + |\varphi^\perp\rangle\langle\varphi^\perp|)$, with $|\varphi\rangle = X_1^{S_2}(\alpha'|\phi^-\rangle + \beta'|\psi^+\rangle)$ and $|\varphi^\perp\rangle = X_1^{S_2}(-\alpha'|\psi^-\rangle + \beta'|\phi^+\rangle)$. Next, measuring qubit 5 in the X basis produces the state $\rho_1 = \frac{1}{2}(|\nu\rangle\langle\nu| + |\nu^\perp\rangle\langle\nu^\perp|)$, with $|\nu\rangle = |\nu^\perp\rangle = X^{S_2}(ZX)^{S_5}Z(\alpha|0\rangle + \beta|1\rangle)$. Thus the final state of qubit 1 is a pure state $\rho_1 = |\nu\rangle\langle\nu|$, from which, if we remove the Pauli operators via feedforward rotations[19], we can recover the encoded qubit and re-encode it for further processing, thereby correcting the loss error. Note that even when there is no loss one can use this method to decode the qubit. A more rigorous description of the recovery procedure using the stabilizer formalism is given in the Methods.

In Figure 3b we show the original Bloch sphere for the ancilla qubit and the recovered sphere after qubit 4 is lost and the recovery is carried out with feedforward rotations. In our graph state qubit 4 is a path qubit and we lose it by incoherently combining the two paths corresponding to its computational basis states. This loss of information about which path photon $s_2$ populates is equivalent to tracing out qubit 4 from the system. Here we have used the four probe states discussed earlier to reconstruct the combined encoding and recovery channel. The recovered Bloch sphere is relatively consistent with the original sphere, corresponding to a process fidelity of $F_p = 0.70 \pm 0.01$, although slightly squeezed in the Z and X directions. This effect can be seen more clearly in the $\chi$ matrix shown in Figure 3c. Here there is a strong component of the identity operation, as expected, but also a non-negligible contribution of a Y operation due to the non-ideal graph resource used in our experiment. The combination of the identity and Y operation gives rise to the squeezing effect seen in the Bloch sphere, which maintains the position of the Y eigenstates, but sends the X and Z eigenstates toward the maximally mixed state $\frac{1}{2}I$. As any state can be written as a combination of these eigenstates, the corresponding components will be affected similarly. The average fidelity for an encoded and recovered qubit is found to be $\bar{F} = 0.82 \pm 0.01$, and the fidelities for the individual probe states are $F_0 = 0.80 \pm 0.01$, $F_1 = 0.77 \pm 0.01$, $F_+ = 0.75 \pm 0.01$ and $F_{+y} = 0.92 \pm 0.01$. In figure 3d we consider qubit 1 as lost and show the Bloch sphere representation of the recovery in Figure 3e, as well as the $\chi$ matrix in Figure 3f. In this case qubit 1 is a polarization qubit and we lose it by removing the PBS at the polarization analysis stage for photon $s_1$, thus combining the two polarizations corresponding to the qubit's computational basis states. We find a process fidelity for the encoding and recovery of $F_p = 0.73 \pm 0.01$. The average fidelity for an encoded and recovered qubit is found to be $\bar{F} = 0.81 \pm 0.01$, and the fidelities for the individual probe states are $F_0 = 0.80 \pm 0.01$, $F_1 = 0.77 \pm 0.01$, $F_+ = 0.78 \pm 0.01$ and $F_{+y} = 0.88 \pm 0.01$. One can see in Figure 3e the recovered Bloch sphere is similar to that of the path qubit loss. However the squeezing is now mainly in the X direction due to the additional presence of a Pauli Z operation, as can be seen more clearly in the $\chi$ matrix shown in Figure 3f.

Finally we check the graph code's ability to detect general quantum errors. To see this note that the logical code states are all common eigenstates of the stabilizer operators $S_1 = Y_1 Z_2 Z_4 Y_5 = K_1 K_5$, $S_2 = Y_1 Z_2 Y_4 Z_5 = K_1 K_4$ and $S_3 = Z_1 Y_2 Y_4 Z_5 = K_4 K_2$, where the $K_i$ are the original graph state stabilizer operators[10,11]. If there is a phase flip Z on any one qubit of the code, as shown in Figure 4a, we can locate the error by measuring all three stabilizers without disturbing the graph code and correct the error as $\langle S_i \rangle = 1$ and $\langle S_i Z_j \rangle = -1$ for a given $j$ and two of the stabilizers. Thus measuring the stabilizers performs the role of syndrome measurements for the graph code. In Figure 4a we show the values of the stabilizers measured in our experiment when there is a Z error on each of the qubits for all the probe states. The experimental values agree well with the theory with all having the correct sign and an error of 0.02 or less. As an arbitrary state can be written as a superposition of the probe states, the results show that any state can be encoded into the code and the error detected. Similar arguments about the stabilizers hold for Y errors, as shown in Figure 4b with the experimental values measured for the probe states. On the other hand, if there is a bit flip X on any one qubit, it can be detected by measuring the stabilizers, but it cannot be located, since an X error anticommutes with all stabilizers: $\langle S_i X_j \rangle = -1$ for a given $j$ and all $i$, as shown by the experimental values in Figure 4c. This is the reason (along with a degeneracy in locating Z and Y errors) why the code can only detect general quantum errors (X, Y or Z) acting on an unknown single qubit, but cannot correct them. If an error is detected via the stabilizers, then the state is discarded and one starts a given quantum protocol again by re-encoding. On the other hand if the location of the error is known, then the type of error (X, Y or Z) can be determined from the pattern of the stabilizer results and the error can be corrected. All expectation values of the stabilizers were found to be consistent with those expected when there was an error occurring on any one of the qubits for all probe states, thus confirming the graph code's ability to detect unknown single-qubit errors and correct known single-qubit errors.

## Discussion

In this work we have reported the experimental demonstration of a graph state code using an all-optical setup to encode quantum information into photons representing the qubits of the code. The experiment was carried out for the smallest graph code capable of correcting up to one general quantum error or a loss error at a known location, or detecting a general quantum error at an unknown location. We showed that the graph state code can be used to correct and detect errors in a photonic setting with the results in close agreement with the theory and limited only by the quality of the initial resource state. Our demonstration and analysis provides a stimulating outlook for several applications of photonic quantum technologies besides the obvious goal of fault-tolerance, for example in communication over lossy channels, lossy interferometry and secret sharing. In general, the versatility of graph codes, such as the one we have realised, can further be increased by generalizing them to CWS codes[60]. As the theory of these codes is the most general theory of QECC at present, the graph encoding we report is amenable to be used with these more general codes. Moreover, the graph code and MBQC techniques we have introduced here can be readily transferred to other promising physical setups, such as ion traps, cavity-QED and superconducting qubits. The next steps will be to design and realize QECC schemes using larger graph states[45,46,47] with enhanced error-correction capabilities[60], and introduce concatenation methods against loss errors[49,48]. Our experimental demonstration and characterization of a four-qubit graph code's performance contributes to the first steps in the direction of full-scale fault-tolerant quantum information processing.

## Methods

**Experimental setup**

The fibre source used was a birefringent PCF similar to that described in refs. 28 and 66. For a pump wavelength of 720 nm launched into the fibre's slow axis, signal-idler pairs are generated on the fast axis at wavelengths of 626 nm and 860 nm, respectively. This is a turning point on the phase-matching curve for the signal wavelength, where the signal spectrum becomes uncorrelated with the pump wavelength, and hence also with the idler spectrum. This means the signal-idler pair are generated almost without spectral correlations in a pure quantum state, and do not require tight spectral filtering to show quantum interference.

To generate entangled pairs from the Sagnac loop source, the fibre axes are rotated at each end. With the fast-axis vertical at the output of the clockwise path, this direction will produce vertical photon-pairs, whereas at the output of the counter-clockwise direction the fast-axis must be horizontal in order to produce horizontal photon-pairs. These orientations also result in the pump light being launched into the correct (slow) axis. Since the pump is always cross polarized from co-propagating photons, it exits the loop from the opposite port, helping to filter it out of the signal and idler channels. A Soleil-Babinet birefringent compensator in the pump beam before the source was used to tune the relative phase between the two terms of the entangled state.

The other PCF source produces horizontally polarized signal photons, which are rotated to diagonal before being fused with the signal from the entangled pair, leaving the three-photon GHZ state. It is necessary to detect the unentangled idler photon from this PCF source in order to herald the signal. The idlers from both sources are filtered with tuneable band pass filters of ~4nm bandwidth to remove Raman emission and other background, while 40nm wide bandpass filters are used for the signals' wavelength which is relatively free of background. All four photons are collected into single-mode fibres. The signals are then relaunched into path-qubit setups, which consist of displaced Sagnac interferometers built around hybrid beamsplitter cubes, with half of the coating a PBS and the other half a 50:50 BS. The photons are split at the PBS side, so their path is correlated with their polarization, and then recombined on the BS side, while the displaced Sagnac configuration gives intrinsic phase stability between the paths. Each path contains a half-wave plate, to carry out the local polarization rotations for state preparation, then a 3mm glass plate, which can be tilted to change the phase and hence the measurement basis.

The signal photons are again collected into single-mode fibres and go to a polarization analysis section. The entangled idler also goes to polarization analysis, but with space for additional optics (a wave plate or polarizer) to be inserted to encode the ancilla qubit state. Polarization analysis consists of a QWP, HWP, then a PBS, with both outputs of the PBS collected into multimode fibres coupled to silicon avalanche photodiodes[67]. The heralding idler goes straight to a detector. The detectors are connected to an eight-channel FPGA[68], which allows all combinations of coincidence to be monitored within a nanosecond-timing window.

**Entanglement witnesses**

For the graph state corresponding to the code resource plus ancilla qubit we use the following entanglement witness on qubits 1, 2, 3, 4 and 5

$$\hat{W} = \frac{9}{4}I - \frac{1}{8}\left(\tilde{X}I\tilde{X}I\tilde{X} + \tilde{X}I\tilde{X}\tilde{X}I + \tilde{X}\tilde{X}I\tilde{X}\tilde{X} + \tilde{X}\tilde{X}III + I\tilde{X}\tilde{X}I\tilde{X} + I\tilde{X}\tilde{X}\tilde{X}I + III\tilde{X}\tilde{X}\right) - \frac{1}{4}(Z\tilde{Y}I\tilde{Y}Z + Z\tilde{Y}\tilde{Y}II + II\tilde{Y}\tilde{Y}Z),$$

where $\tilde{O}$ corresponds to measurements in the $O$ basis with the eigenstates swapped. This is a locally rotated version of the witness given in ref. 69 for a five-qubit linear cluster state and takes into account the local complementation operations described in Figure 1b of the main text.

For the box cluster we use the following entanglement witness on qubits 1, 2, 4 and 5

$$\widehat{W} = 2I - \frac{1}{2}(ZI\tilde{X}\tilde{X} + Z\tilde{Z}\tilde{X}I + II\tilde{X}\tilde{X} + \tilde{X}\tilde{X}IZ + \tilde{X}\tilde{X}II + I\tilde{X}\tilde{Z}Z),$$

which is a locally rotated version of the one given in ref. 69 for a four-qubit linear cluster state and takes into account the local complementation operations needed to rotate it into a box-cluster.

For the rotated GHZ state we use the following entanglement witness on qubits 1, 2, 4 and 5

$$\widehat{W} = \frac{7}{4}I - ZZ\tilde{Z}\tilde{Z} - \frac{1}{4}(XXII + XIXI + XIIX + IXXI + IXIX + IIXX + XXXX),$$

which is a locally rotated version of the one given in ref. 69.

For the maximally entangled qubit pairs in the logical encoding of the probe state $|+_y\rangle$ we use the following entanglement witness on qubit pair (1,2) and pair (4,5)

$$\widehat{W} = I - \tilde{Y}Z - XX$$

which is a locally rotated version of the one given in ref. 69 for a two-qubit linear cluster state.

**Stabilizer picture of the graph code**

The stabilizer description of QECC is a compact and powerful way to gain insight on the symmetries of quantum codes. A different way of writing the original state of the ancilla qubit 3 is $|\psi\rangle_3 = \frac{1}{2}(I + e_x X_3 + e_y Y_3 + e_z Z_3)|\psi\rangle_3$, where $e_x^2 + e_y^2 + e_z^2 = 1$. In order to see how this description is equivalent to the one introduced in the Results section, note that $|\psi\rangle_3 = \alpha |0\rangle_3 + \beta |1\rangle_3 = U |0\rangle_3$ for some unitary operation $U$. The projector will transform accordingly, i.e. $|\psi\rangle_3\langle\psi| = U|0\rangle_3\langle 0|U^\dagger = \frac{1}{2}(I + UZ_3 U^\dagger) = \frac{1}{2}(I + e_x X_3 + e_y Y_3 + e_z Z_3)$, since the Pauli matrices, together with the identity, form a basis for all single-qubit density matrices. Specifically we have that, for $\alpha = \cos\frac{\theta}{2}$ and $\beta = e^{i\varphi} \sin\frac{\theta}{2}$, the correspondence $e_x = \sin\theta \cos\varphi$, $e_y = \sin\theta \sin\varphi$ and $e_z = \cos\theta$.

The four qubit graph code [[4,1,2]] is the common eigenspace of the stabilizer operators $S_1 = Y_1 Z_2 Z_4 Y_5 = K_1 K_5$, $S_2 = Y_1 Z_2 Y_4 Z_5 = K_1 K_4$ and $S_3 = Z_1 Y_2 Y_4 Z_5 = K_4 K_2$, where the $K_i = X_i \otimes_{j \in N(i)} Z_j$ are the original box cluster state stabilizer operators[10,11]. We have chosen $\overline{X} = Z_1 Z_2 X_4 I_5$ and $\overline{Z} = Z_1 Z_2 Z_4 Z_5$ to be the logical Pauli operators acting on the codespace, respectively. Note that this choice is independent from the labelling. Encoding information can be seen as *expanding* the operators acting on the ancilla qubit into the four-qubit box cluster state plus ancilla. For simplicity, we fix $e_y = 0$, and restrict the logical state to be in the X-Z equator of the Bloch sphere. After tracking how the X and Z operators are expanded, we then find the expansion of Y = $i$XZ and remove the restriction. Note that the controlled-phase gate acts like $C_Z^{ij}(I_i X_j) C_Z^{ij\dagger} = Z_i X_j$ and $C_Z^{ij}(X_i I_j) C_Z^{ij\dagger} = X_i Z_j$. Applying the operation $C_Z^T = C_Z^{13} C_Z^{23} C_Z^{43} C_Z^{53}$ to the qubits of the four-qubit box cluster and an ancilla qubit to make the initial five-qubit graph state resource (code plus ancilla) will change the shape of the logical ancilla operators as:

$$\overline{X}_e = C_Z^T X_3 C_Z^{T\dagger} = \begin{matrix} Z_2 & Z_4 \\ X_3 & \\ Z_5 & Z_1 \end{matrix}$$

$$\overline{Z}_e = C_Z^T Z_3 C_Z^{T\dagger} = \begin{matrix} I_2 & I_4 \\ Z_3 & \\ I_5 & I_1 \end{matrix}$$

We can reshape these expanded logical operators by multiplying them by expanded versions of operators $O$ for which the box cluster is an eigenstate, *i.e.* $\overline{X}'_e \equiv \overline{X}_e \tilde{O}$, where the operators $\tilde{O} = C_Z^T O C_Z^{T\dagger}$:

$$\overline{X}'_e = \overline{X}_e \tilde{S}_1 = \begin{matrix} Z_2 & \\ X_3 & \\ Z_5 & \end{matrix} \begin{matrix} Z_4 & Z_2 \\ & \\ Z_1 & Y_5 \end{matrix} \cdot \begin{matrix} & \\ I_3 & \\ & Y_1 \end{matrix} = \begin{matrix} Z_4 & I_2 \\ & X_3 \\ & X_5 \end{matrix} \begin{matrix} I_4 \\ \\ X_1 \end{matrix}$$

and

$$\overline{Z}'_e = \overline{Z}_e \tilde{K}_5 = \begin{matrix} I_2 & \\ Z_3 & \\ I_5 & \end{matrix} \begin{matrix} I_4 & Z_2 \\ & \\ I_1 & X_5 \end{matrix} \cdot \begin{matrix} I_4 & \\ Z_3 & \\ & Z_1 \end{matrix} \begin{matrix} Z_2 & \\ & I_3 \\ X_5 & \end{matrix} \begin{matrix} I_4 \\ \\ Z_1 \end{matrix}$$

where the operator $\widetilde{K}_5 = Z_1 Z_2 I_4 X_5$ is a cluster state stabilizer. Since the expanded logical operators do not have support on qubit 4, measuring this qubit will not be needed to decode the information, and it can thus be lost. Qubit 3 will be measured in the X basis, which leaves the four remaining qubits in the state $|\psi_L\rangle = \frac{1}{2}(I + e_z \bar{X} + e_x \bar{Z}) |\psi_L\rangle$. It is straightforward to see that the encoding operation entangles ancilla qubit 3 with the qubits of the code, and its measurement in the X basis effectively teleports the information into the codespace, after an application of a Hadamard operation (note the unit vectors $e_x$ and $e_z$ are swapped in the encoded state). We can then find the logical Y operator using the relation $\bar{Y} = i\overline{XZ}$ to generalize the result. Of the qubits in the graph code, one can see from the form of the logical operators that we need to measure qubits 2 and 5 in the Z and X basis, respectively. That will leave qubit 1 in the state $|\psi\rangle$, modulo some known Pauli corrections. This method constitutes a generalization to logical subspaces of the task for propagating information through a resource state in MBQC. The above method also illustrates how decoding can be achieved.

**Acknowledgements** We thank T. Rudolph for theory discussions, and A. McMillan and R. Nock for experimental discussions. This work was supported by the UK's Engineering and Physical Sciences Research Council, ERC grant 247462 QUOWSS, the National Research Foundation and Ministry of Education, Singapore, the Leverhulme Trust, the HIPERCOM (2011-CHRI-006) project and the Ville de Paris Emergences program, project CiQWii.

**Competing interests statement** The authors declare that they have no competing financial interests.

**Correspondences** and requests for materials should be addressed to

M. S. Tame (markstame@gmail.com) or J. G. Rarity (j.rarity@bristol.ac.uk).

**Figure 1. Experimental setup. a**, Setup used to generate the graph state resource consisting of the four-qubit graph code plus ancilla qubit. Two photonic crystal fibre sources are pumped using a Ti:Sapphire laser producing picosecond pulses at 720 nm. The first source produces a pair of photons in the state $|H\rangle_{i_1}|H\rangle_{s_1}$ and the second produces photons in the state $\frac{1}{\sqrt{2}}(|H\rangle_{i_2}|H\rangle_{s_2}+|V\rangle_{i_2}|V\rangle_{s_2})$. The signal photons from the first pair are rotated to the state $|+\rangle$ using a half-wave plate (HWP) and both signal photons are then fused using a polarizing beamsplitter. The polarizations of the signal photons are then rotated using HWPs to form the three-qubit linear cluster state $\frac{1}{\sqrt{2}}(|+\rangle_{s_1}|H\rangle_{i_2}|+\rangle_{s_2}+|-\rangle_{s_1}|V\rangle_{i_2}|-\rangle_{s_2})$, where the first idler photon is used as a trigger to verify a four-fold coincidence signifying the generation of the state. The path degree of freedom of the signal photons is then used to expand the resource to a five-qubit linear cluster state using a Sagnac interferometer, as shown in the dashed boxes and explained in the main text. **b**, Five-qubit linear cluster state and local complementation steps (LC$_1$ and LC$_2$) to generate the graph code plus ancilla qubit. Here, the vertices correspond to qubits initialized in the state $|+\rangle$ and edges correspond to controlled-phase gates, $C_Z = \mathrm{diag}(1,1,1,-1)$, applied to the qubits. The LC operations are performed using half-wave plates, quarter-wave plates and phase shifters in the relevant photon modes and correspond to LC$_1$ = A$_1$B$_2$(AA)$_3$B$_4$A$_5$ and LC$_2$ = A$_1$A$_2$B$_3$A$_4$A$_5$, where A = $\sqrt{-iZ}$ and B = $\sqrt{-iX}$. In the steps shown in the figure, the operation A (B) is depicted as a dashed (solid) outline around the qubit. **c**, Expectation values of the operators used to verify genuine multipartite entanglement in the graph state and obtain a lower bound on the fidelity. Here $\tilde{O}$ corresponds to measurements in the $O$ basis with the eigenstates swapped. The ideal values correspond to the dashed line.

**Figure 2. Graph code. a**, Encoding logical states. In order to encode the state of the ancilla qubit into the graph it should be measured in the X basis. This propagates the information into the graph code while at the same time applying a Hadamard operation. Thus the ancilla state is encoded in the Hadamard basis. **b**, Logical density matrices for the four different probe states $|0\rangle$, $|1\rangle$, $|+\rangle$ and $|+_y\rangle$ once propagated into the code. These are calculated from the expectation values of the joint four-qubit logical operators $\overline{X}$, $\overline{Y}$ and $\overline{Z}$. **c**, Encoding as a channel. Here the Bloch sphere transformation is shown for the encoding of arbitrary ancilla qubits (points on the surface of the sphere) into the code. Note that a Hadamard operation has been performed on the qubit, corresponding to a rotation of 180 degrees about the X-Z plane.

**Figure 3. Loss tolerance. a**, General scenario of loss tolerance for the four-qubit graph code. Here any one of the four qubits may be lost. In the first case, qubit 4 has been lost by combining the two paths corresponding to the computational basis of the qubit. The encoded qubit can be recovered on qubit 1 using the measurements and results of the remaining qubits 2 and 5 as described in the main text. **b**, Path qubit lost with the recovery treated as a channel. Here the Bloch sphere representation is used to show the original qubit states and the recovered qubit states. **c**, The $\chi$ matrix representation of the channel, showing the real part (left) and imaginary part (right). Ideally the $\chi$ matrix has only one component, the entry $II$, corresponding to the identity operation. **d**, In the second case, qubit 1 has been lost by combining the two polarizations corresponding to the computational basis of the qubit. The encoded qubit is recovered on qubit 5 using the measurements and results of the remaining qubits 2 and 4. **e**, Polarization qubit loss with the recovery treated as a channel. Here the Bloch sphere representation shows the original qubit states and the recovered qubit states. **f**, The $\chi$ matrix

representation of the channel, showing the real part (left) and imaginary part (right).

**Figure 4. Error detection. a**, Z errors on one of the qubits of the code flips the sign of the expectation value of one or two of the stabilizer (syndrome) operators $S_1$, $S_2$ and $S_3$, as can be seen in the tables showing the experimental values for the four probe states. The values range from 0.66 to 0.79 in magnitude. The syndrome operators correspond to joint measurements, thus they can in principle be measured without disturbing the state. If no error has occurred the code can continue to be used. If an error has occurred then it will be detected and the ancilla can be encoded again to allow the continuation of a given protocol. If the error is known to be a Z Pauli operation then its location can be detected and corrected. If it is not, the ancilla must be re-encoded to allow the continuation of a given protocol. **b**, Y errors on one of the qubits of the code also flips the sign of the expectation value of one or two of the syndrome operators. If the error is known to be a Y Pauli operation then its location can be detected and corrected. If not, the ancilla can again be re-encoded. **c**, X errors on one of the qubits of the code flips the sign of the expectation value of all the syndrome operators. Note that if the location of the error is known, then the type of error can be inferred from the pattern of the expectation values of the syndrome operators and the error can be corrected.

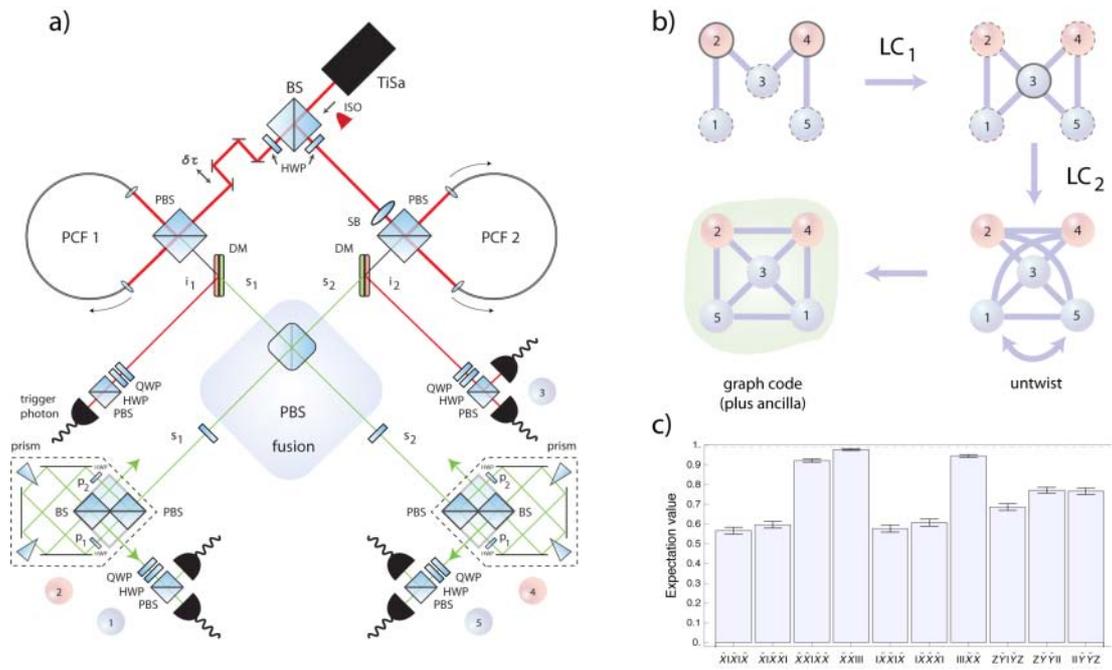

**Figure 1**

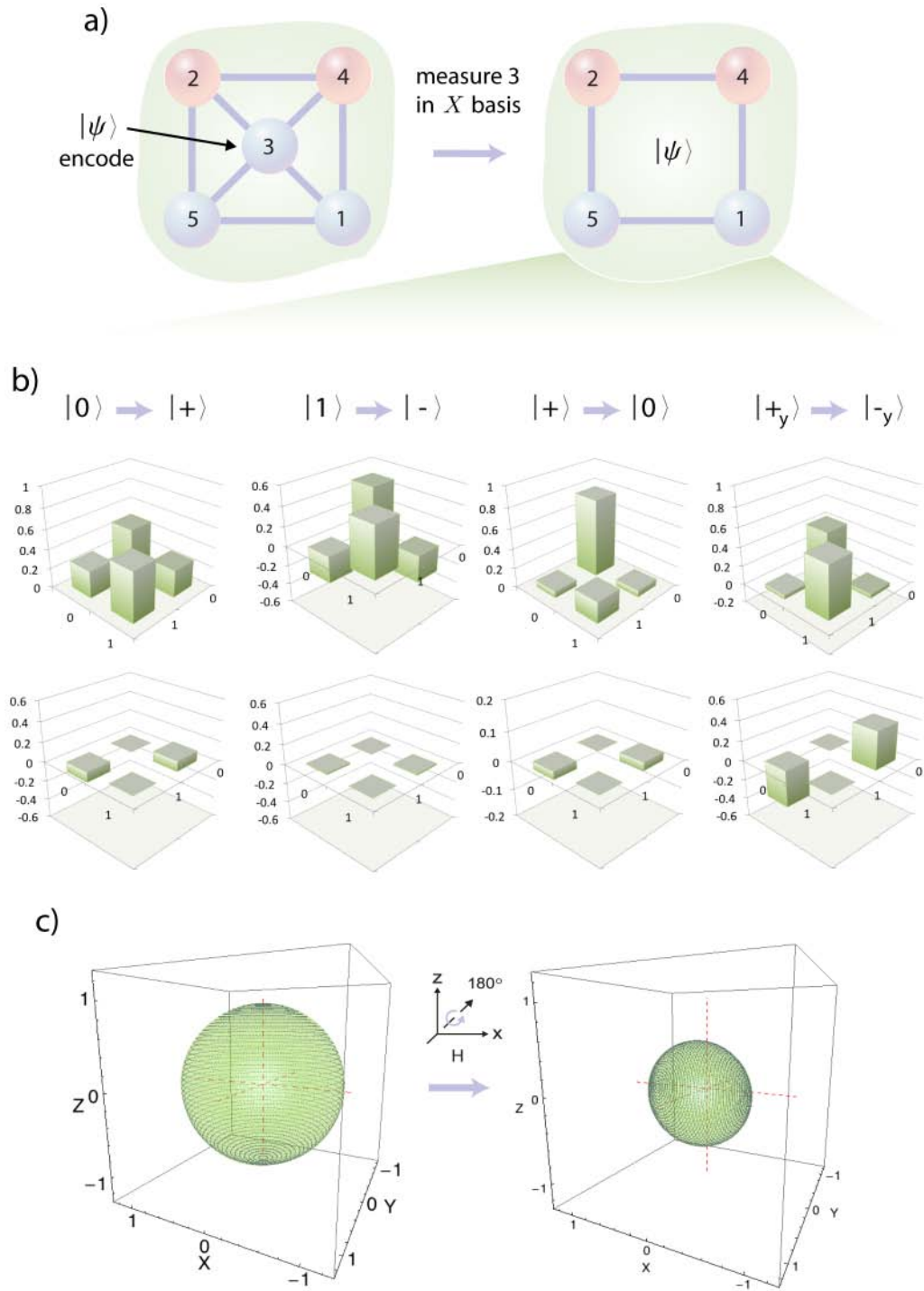

**Figure 2**

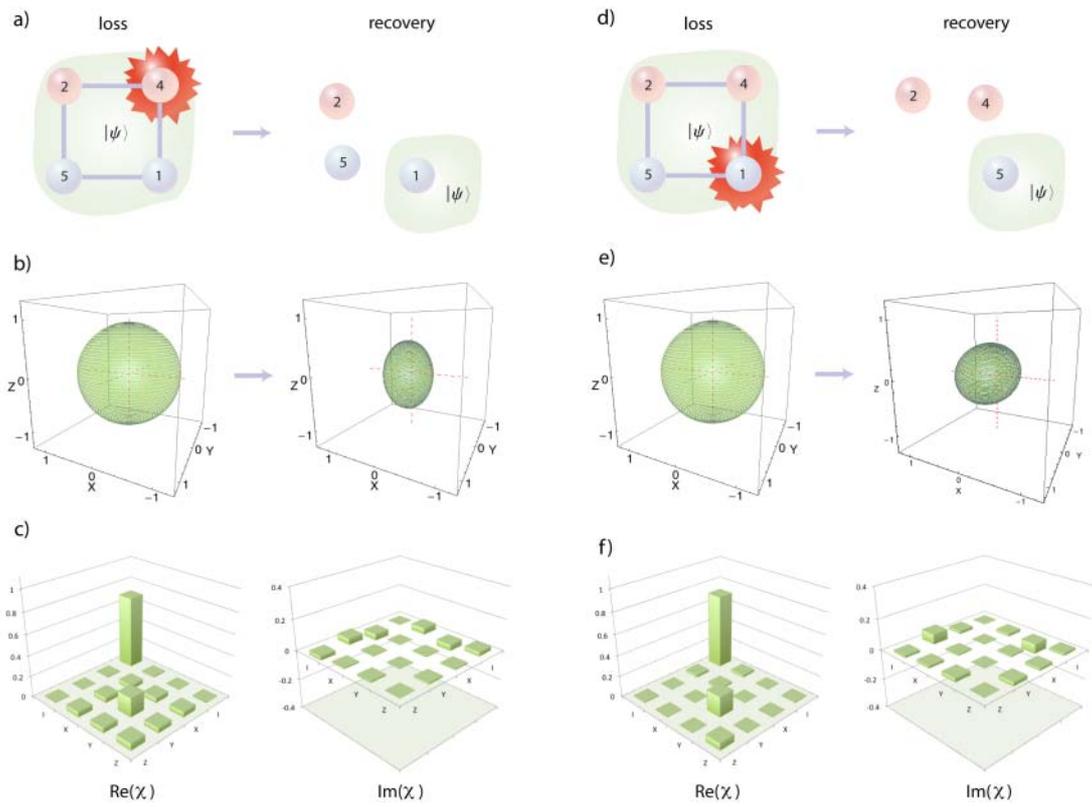

**Figure 3**

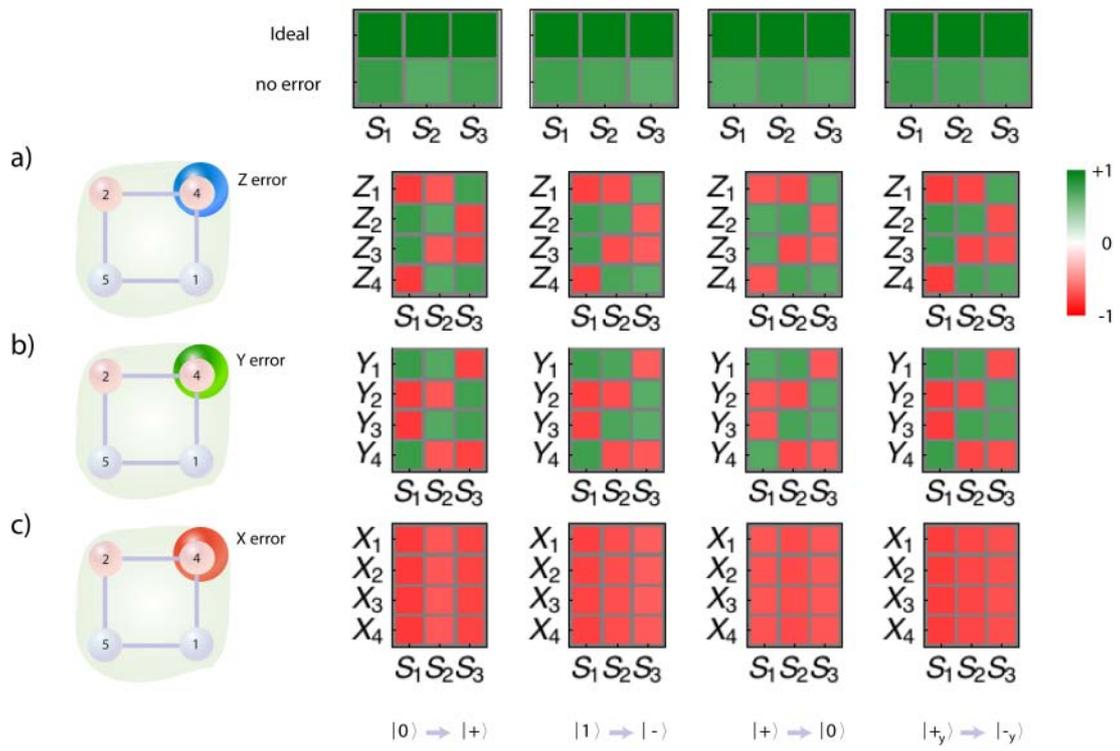

**Figure 4**